\documentclass[sigconf]{acmart}

\acmConference[EASE 2026]{The 30th International Conference on Evaluation
and Assessment in Software Engineering}{9–12 June, 2026}{Glasgow, Scotland,
United Kingdom}

\setcitestyle{numbers,sort&compress}
\usepackage{algorithmic}
\usepackage{graphicx}
\usepackage{textcomp}
\usepackage{xcolor}
\usepackage{multirow}
\usepackage{booktabs}
\usepackage{tcolorbox} 
\usepackage{caption}
\usepackage[table]{xcolor}
\usepackage{hyperref}
\usepackage{hyperref}

\usepackage{enumitem}

\hypersetup{
    colorlinks=true,
    linkcolor=blue,
    citecolor=blue,
    urlcolor=blue,
}
\definecolor{highlight}{RGB}{200,225,200}   
\definecolor{border}{RGB}{25,75,25}   
\definecolor{best}{RGB}{198,239,206}   
\definecolor{worst}{RGB}{255,199,206}  
\def\BibTeX{{\rm B\kern-.05em{\sc i\kern-.025em b}\kern-.08em
    T\kern-.1667em\lower.7ex\hbox{E}\kern-.125emX}}

\newcommand{\strategy}[1]{\texttt{#1}}

\begin{document}

\title{Context Matters: Evaluating Context Strategies for Automated ADR Generation Using LLMs}

\author{Aviral Gupta}
\email{aviral.gupta@research.iiit.ac.in}
\affiliation{%
  \institution{Software Engineering Research Centre, IIIT Hyderabad}
  \country{India}
}
\author{Rudra Dhar}
\email{rudra.dhar@research.iiit.ac.in}
\affiliation{%
  \institution{Software Engineering Research Centre, IIIT Hyderabad}
  \country{India}
}
\author{Daniel Feitosa}
\orcid{0000-0001-9371-232X}
\email{d.feitosa@rug.nl}
\affiliation{%
  \institution{University of Groningen}
  \country{The Netherlands}
}
\author{Karthik Vaidhyanathan}
\email{karthik.vaidhyanathan@iiit.ac.in}
\affiliation{%
  \institution{Software Engineering Research Centre, IIIT Hyderabad}
  \country{India}
}

\renewcommand{\shortauthors}{Gupta et al.}

\setlength{\floatsep}{1pt plus 1pt minus 1pt}
\setlength{\textfloatsep}{1pt plus 1pt minus 1pt}
\setlength{\intextsep}{1pt plus 1pt minus 1pt}
\setlength{\dblfloatsep}{1pt plus 1pt minus 1pt}
\setlength{\dbltextfloatsep}{1pt plus 1pt minus 1pt}
\setlength{\abovecaptionskip}{2pt}
\setlength{\belowcaptionskip}{3pt}

\begin{abstract}
Architecture Decision Records (ADRs) play a critical role in preserving the rationale behind system design, yet their creation and maintenance are often neglected due to the associated authoring overhead. This paper investigates whether Large Language Models (LLMs) can mitigate this burden and, more importantly, how different strategies for presenting historical ADRs as context influence generation quality. We curate and validate a large corpus of sequential ADRs drawn from 750 open-source repositories and systematically evaluate five context selection strategies (no context, All-history, First-K, Last-K, and RAFG) across multiple model families. Our results show that context-aware prompting substantially improves ADR generation fidelity, with a small recency window (typically 3–5 prior records) providing the best balance between quality and efficiency. Retrieval-based context selection yields marginal gains primarily in non-sequential or cross-cutting decision scenarios, while offering no statistically significant advantage in typical linear ADR workflows. Overall, our findings demonstrate that context engineering, rather than model scale alone, is the dominant factor in effective ADR automation, and we outline practical defaults for tool builders along with targeted retrieval fallbacks for complex architectural settings.
\end{abstract}



\maketitle
    
\vspace{-2pt}
\section{Introduction}

Software architecture is not merely a static structure but a continuous stream of design decisions made to satisfy evolving requirements and constraints~\cite{kruchten2009capilla}. To manage this complexity, Architecture Decision Records (ADRs) have emerged as a \textit{de facto} standard for lightweight Architectural Knowledge Management (AKM)~\cite{nygard2011,Zimmermann2011,Capilla2016}. By capturing the context, decision, and consequences of architectural changes, ADRs facilitate team onboarding, prevent knowledge vaporization, and maintain traceability in long-lived systems.

Despite their recognized value, the practice of writing ADRs often suffers from a misalignment of incentives: the cost of documentation is paid immediately by the architect, while the benefits are reaped primarily by future maintainers. Consequently, ADR adoption in practice is inconsistent. A recent large-scale study of open-source repositories revealed that while many projects initialize ADR collections, a significant portion abandons the practice shortly thereafter, leaving decision logs incomplete or outdated~\cite{buchgeher2023using}. This ``documentation debt'' hinders architectural understanding and increases the risk of architectural drift.

The rapid ascendancy of Large Language Models (LLMs) presents a compelling opportunity to mitigate this effort. LLMs have demonstrated significant impact across numerous Software Engineering (SE) tasks, including code summarization, requirements engineering, and text generation~\cite{Grmez2024, hou2024llm4se}, making them theoretically well-suited for automating the drafting of ADRs. Preliminary studies have explored this potential, showing that models like GPT-4 can generate plausible decision summaries when provided with specific prompts~\cite{dhar2024can}. However, existing research has largely treated ADR generation as an isolated task—generating a single decision based on immediate code changes or a narrow context window~\cite{dhar2024exploratory}.

In the landscape of modern LLM applications, Retrieval-Augmented Generation (RAG) has emerged as a critical technique to enhance model capabilities by providing external knowledge. However, practitioners face significant challenges regarding fundamental engineering trade-offs: deciding how much information to retrieve and how to integrate that knowledge effectively without introducing noise~\cite{zhao2025dddrag}. These trade-offs are particularly acute in software architecture, where providing too little context leads to generic decisions, while too much context may exceed token limits or distract the model with irrelevant historical data.

This isolated approach overlooks a fundamental property of software architecture: \textit{path dependency}. An architectural decision is rarely made in a vacuum; it is constrained by the history of prior decisions, the established technology stack, and previously accepted trade-offs. We posit that the quality of an ADR generation depends critically not just on the capability of the underlying model, but on the \textbf{context engineering} strategy used to retrieve this historical narrative~\cite{shi2025context}.

In this paper, we present a comprehensive empirical study evaluating how different context-provision strategies influence the quality of LLM-generated ADRs. Unlike previous works that focus primarily on comparing models, we rigorously control for the architectural context provided to those models. To support this investigation, we utilized a large-scale corpus of ADRs originally identified in an MSR study by Buchgeher et al.~\cite{buchgeher2023using}. Building on this foundation, we performed extensive manual filtering and sequence verification across 750 repositories to ensure chronological integrity, thereby transforming the raw data into a validated sequential dataset that simulates real-world architectural evolution.

Specific contributions of this paper are:
\begin{itemize}[nosep, leftmargin=*]
    \item to provide a curated corpus of ADRs that preserves chronological ordering, enabling valid ``next-step'' generation experiments.
    \item to study the influence of decision history in prompt for generating an ADR through LLM
    \item to quantify the performance impact of four distinct context strategies: \textit{All-History}, \textit{First-K} (foundational context), \textit{Last-K} (recent context), and \textit{RAFG} (semantic relevance).
\end{itemize}

The remainder of this paper is organized as follows. Section \ref{sec:background}  reviews prior work and situates our study in the relevant literature. Next, Section \ref{sec:methods} describes our experimental methodology, followed by Section \ref{sec:results} presenting our empirical findings. Section \ref{sec:discussion} interprets these results, outlines implications, and addresses validity threats.  We then review additional related work in Section \ref{sec:relatedWork}, and conclude with a summary of contributions and directions for future research in Section \ref{sec:conclusions}. Finally, we provide information on data availability in Section \ref{sec:data}.

\vspace{-2pt}
\section{Background}\label{sec:background}

This work sits at the intersection of Architectural Knowledge Management, Generative AI, and Empirical Software Engineering. In this section, we outline the foundation of ADR in practice, and recent studies that also touch upon the mentioned intersection, motivating our investigation into context-aware ADR generation.

\subsection{Architecture Decision Records in Practice}
Software architecture can be fundamentally defined as a set of architectural design decisions~\cite{Jansen2005soft}. To capture and manage these decisions, ADRs were popularized by Nygard as a lightweight, version-controlled method to capture the ``why'' behind software architecture. Despite their theoretical benefits for knowledge management, widespread industrial adoption remains challenging. Buchgeher et al.~\cite{buchgeher2023using} conducted a large-scale Mining Software Repositories (MSR) study analyzing over 900 repositories on GitHub. Their findings reveal a critical ``adoption gap'': while the number of projects initializing ADRs is increasing, approximately 50\% of these repositories contain fewer than five records, suggesting a high rate of abandonment after an initial pilot phase. Furthermore, they observed that sustained ADR maintenance is typically a collaborative effort rather than the responsibility of a single architect. This empirical evidence demonstrates the need for automated tooling to reduce the friction of creating and maintaining documentation, which motivates our investigation into LLM-driven generation.

\begin{figure}[ht]
\centering
\begin{tcolorbox}[
    colback=black!5!white,
    colframe=black!60!black,
    boxrule=0.5pt,
    arc=2pt,
    left=6pt,
    right=6pt,
    top=6pt,
    bottom=6pt
]
\textbf{ADR 01: Use Jest as the Testing Framework}

\vspace{0.5em}
\textbf{Status:} Accepted

\vspace{0.5em}
\textbf{Context:}  
We want a testing framework with strong support for TypeScript and Node.js. Jest provides fast execution, rich mocking capabilities, and a mature ecosystem.

\vspace{0.5em}
\textbf{Decision:}  
We will use Jest as the primary testing framework.

\vspace{0.5em}
\textbf{Consequences:}
\begin{itemize}
    \item All tests will be written using Jest.
    \item Jest's built-in mocking and TypeScript support can be leveraged.
\end{itemize}
\end{tcolorbox}
\caption{Example of an Architecture Decision Record (ADR)}
\label{fig:ADR}
\end{figure}

\subsection{Context Engineering and Retrieval-Augmented Generation}

A recurring finding in the literature is that generative quality is bounded by the relevance of the provided context. In this direction, Yu et al.~\cite{yu2024droidcoder} found that `context-enriched' Retrieval-Augmented Generation (RAG) significantly outperforms standard retrieval by including semantic relationships rather than just keyword matches. Similarly, Fuch{\ss} et al.~\cite{fuchs2025lissa} addressed the specific challenge of traceability in documentation through the LiSSA framework. Using RAG, they demonstrated that recovering semantic links between documentation and implementation artifacts is essential for keeping documentation ``live.'' Their work supports the hypothesis that RAG-based context strategies may outperform rigid chronological strategies by filtering out noise and focusing the LLM on architecturally relevant history.

In this study we use Retrieval-Augmented Few-shot Generation (\strategy{RAFG})~\cite{izacard2022rafg}, which is a combination of few-shot prompting and RAG. Rather than relying on a static, predefined set of few-shot exemplars, this approach retrieves contextually similar examples from a structured knowledge base, such as a vector database (VDB).

\vspace{-2pt}
\section{Study Design and Execution}\label{sec:methods}

\begin{figure}[ht]
    \centering
    \includegraphics[width=1\linewidth]{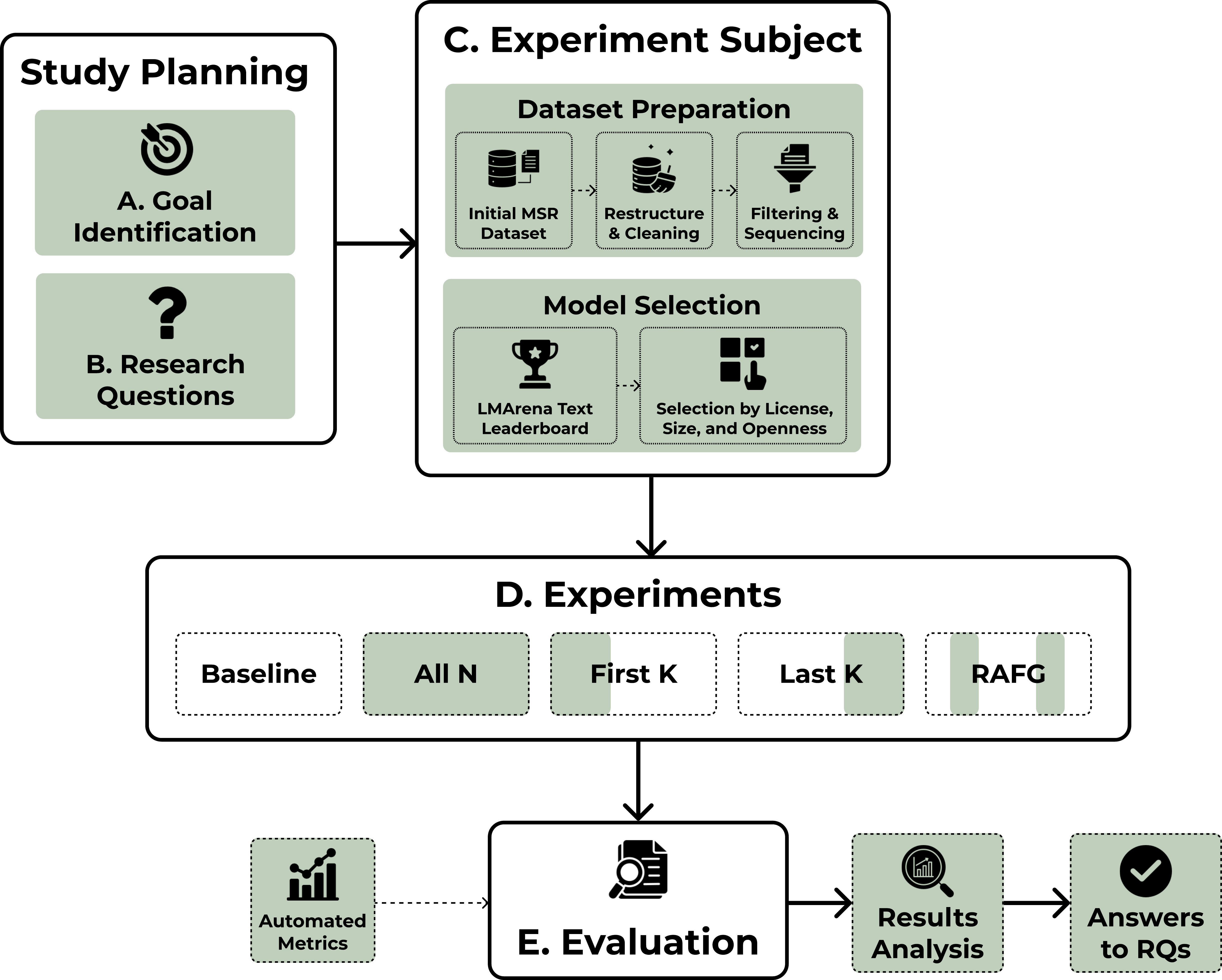}
    \caption{Study Design}
    \label{fig:stuDesign}
\end{figure}

This section presents our empirical methodology. We describe the study goal (Section~\ref{subsec:goal}), derive research questions (Section~\ref{subsec:rq}), outline the study subjects including the construction of the data set and the selection of models (Section~\ref{subsec:expSub}), detail the experimental procedure and context-building strategies (Section~\ref{subsec:expProc}), and conclude with the evaluation metrics used to assess the quality of ADR (Section~\ref{subsec:metrics}). Figure~\ref{fig:stuDesign} provides an overview of the full experiment workflow.

\subsection{Goal}\label{subsec:goal}
The primary goal of this empirical study is to evaluate how different context-provision strategies affect the ability of LLMs to generate an ADR when provided only with an ADR title as the target prompt. More formally,
by utilizing the Goal-Question-Metric approach~\cite{basili1994gqm}, the objective of this study can be described as follows:

\noindent
\textbf{Analyze} the effectiveness of context \\
\textbf{For the purpose of} generating Architecture Decision Records \\
\textbf{With respect to} different context selection strategies \\
\textbf{From the viewpoint of} Software Architects \\
\textbf{In the context of} using Generative AI for AKM.

\subsection{Research Questions}\label{subsec:rq}\
To achieve this goal, we structured the study around the following research questions:

\textbf{$\mathbf{RQ_1}$ \textit{How does architectural decision history influence the quality of LLM-generated ADRs?}}

This research question examines whether providing prior architectural decision history improves the quality of ADRs generated by LLMs compared to a no-context baseline (where no prior decisions are given to the LLM). Since architectural decisions are path-dependent and constrained by previously accepted trade-offs and technologies, this RQ evaluates whether historical grounding enables models to generate ADRs that better reflect realistic architectural rationale. The outcome clarifies whether ADR generation should be treated as a context-aware task rather than an isolated text generation problem.

\textbf{$\mathbf{RQ_2}$ \textit{Is the complete architectural decision history necessary for effective ADR generation, or can a strategically selected subset achieve comparable performance?}}

This question explores the trade-off between completeness and efficiency in context provisioning. While including the full ADR history may theoretically offer the most comprehensive architectural grounding, it introduces scalability challenges such as increased token usage, noise from obsolete decisions, and higher computational cost. We therefore examine if selective context strategies, e.g., using only foundational decisions (First-K) or recent decisions (Last-K), can approximate or match the quality of full-history prompting. The goal is to identify whether there exists a minimal yet sufficient context window that captures the most architecturally salient information needed for accurate ADR generation.

\textbf{$\mathbf{RQ_3}$ \textit{How does RAFG compare to using a chronological sequence of prior ADRs as context?}}

This research question examines whether retrieval-based, semantically relevant context selection offers advantages over chronological context strategies for ADR generation. While chronological approaches emphasize temporal proximity, RAFG prioritizes conceptual relevance and may surface architecturally pertinent but temporally distant decisions. This RQ assesses whether the added complexity of retrieval-based context selection yields meaningful improvements, particularly in non-linear or cross-cutting architectural scenarios.

\subsection{Experiment Subject}\label{subsec:expSub}\
    
\subsubsection{\textbf{Dataset Preparation}}
We curated a high-quality dataset of ADR sequences by refining the corpus originally published by Buchgeher et al.~\cite{buchgeher2023using}. While the original study established a foundational collection, our preliminary analysis revealed significant noise and structural inconsistencies that required further pre-processing. Specifically, we identified:
\begin{itemize}
    \item Structural Heterogeneity: Inconsistent directory nesting and naming conventions across repositories.
    \item Temporal Discontinuity: Missing or non-sequential ADR indices that obscured the decision-making timeline.
    \item Artifact Noise: The prevalence of boilerplate templates, empty stubs, or non-architectural documentation.
\end{itemize}

To ensure the integrity of the sequential data, we performed a multi-stage manual filtering and validation process:
\begin{enumerate}[nosep, leftmargin=*]
    \item Repository Structure Audit: We manually inspected each repository to isolate dedicated ADR directories and distinguish genuine architectural records from general documentation.
    \item Chronological Sequencing: We reconstructed the chronological order of ADRs by reconciling numeric prefixes, file-system metadata, and version control history.
    \item Metadata Indexing: Finally, each repository was cataloged with descriptive metadata, including language composition and popularity metrics (stars, forks, and open issues), to provide context for the architectural maturity of the projects.
\end{enumerate}

The dataset cleaning and validation process was performed by the first author using predefined, rule-based criteria (e.g., directory structure, naming conventions, and sequential consistency). Manual intervention was limited to resolving structural ambiguities rather than subjective content evaluation, minimizing individual bias. All procedures are reproducible via the replication package\footnote{ Replication Package: https://zenodo.org/records/18370195}.

The resulting dataset comprises over 4,500 validated ADRs spanning 750 repositories, representing one of the largest curated sequential ADR corpora to date (to the best of our knowledge). This curated dataset forms the foundation for all experiments in this study.

\begin{figure}[ht]
    \centering
    \includegraphics[width=1\linewidth]{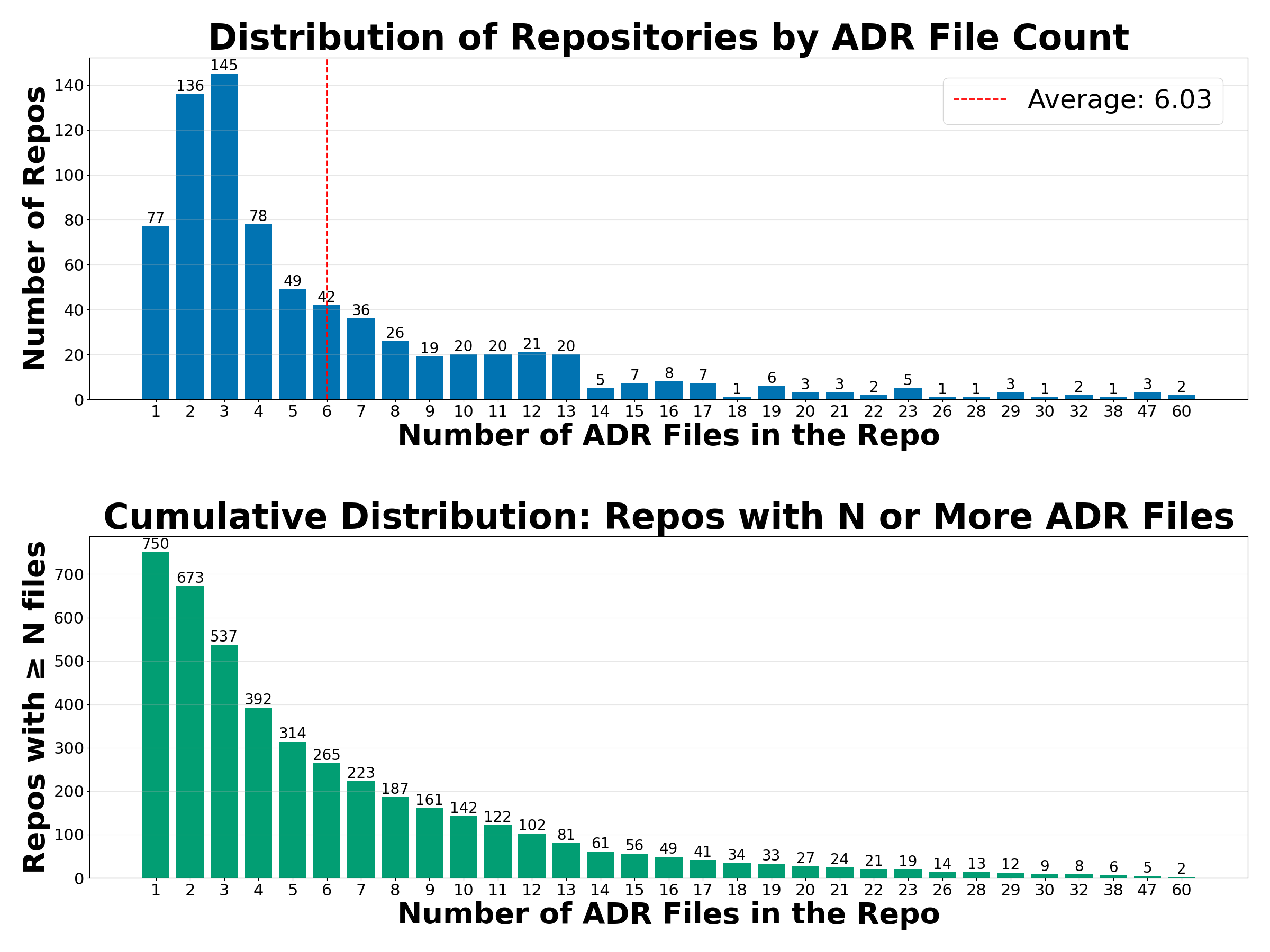}
    \caption{ADR Frequency Distribution}
    \label{fig:adr_count}
\end{figure}

The distribution of ADRs per repository shows a significant right-skew, as illustrated in Figure~\ref{fig:adr_count}. While the average repository contains approximately 6.03 ADRs, the distribution is heavily concentrated in the 1 to 4 range. This indicates that while many projects adopt ADRs, only a small subset (of less than 2\% of the sample) maintains a longitudinal history exceeding 25 records. This distribution highlights the challenge of capturing long-term architectural evolution in open-source software.

\subsubsection{\textbf{Large Language Models (LLMs)}}
To ensure a comprehensive evaluation, we selected a heterogeneous suite of models representing a broad spectrum of computational scales, licensing, and architectural paradigms. Selection was guided by empirical performance rankings on the LMArena Text Leaderboards\footnote{ Web archive snapshot of LMArena Text Leaderboard taken October 15, 2025:  https://web.archive.org/web/20251015124736/https://lmarena.ai/leaderboard/text)} as well as practical considerations regarding accessibility and operational cost. The following models were selected for evaluation:
\begin{enumerate}[nosep, leftmargin=*]
    \item Proprietary State-of-the-Art (SOTA): \textit{Gemini-2.5-Pro}~\cite{gemini25} is a proprietary closed-source model offering strong reasoning capabilities and a good performance-to-cost profile. Gemini represents a large-scale multimodal transformer architecture.
    \item Best Performing Open-Source Model: \textit{Qwen3-235B-A22B-Instruct-2507}~\cite{qwen3} is an instruction-tuned, open-source model that reflects the current frontier of publicly available transformer-based LLMs.
    \item Open-Source Model with Alternative Architecture: \textit{GLM-4.6}~\cite{glm46} introduces a graph-augmented modeling approach that departs from strictly sequential transformer attention. Its inclusion expands architectural diversity and enables comparison between conventional and graph-enhanced language models.
    \item Locally Executable Lightweight Model: \textit{Gemma-3-4B-it}~\cite{gemma3} is a compact, instruction-tuned model optimized for local deployment, suitable for organizations requiring strict data-governance controls or low-resource execution environments.
\end{enumerate}

\subsection{Experiment Procedure}\label{subsec:expProc}\

\begin{figure}[ht]
    \centering
    \includegraphics[width=0.9\linewidth]{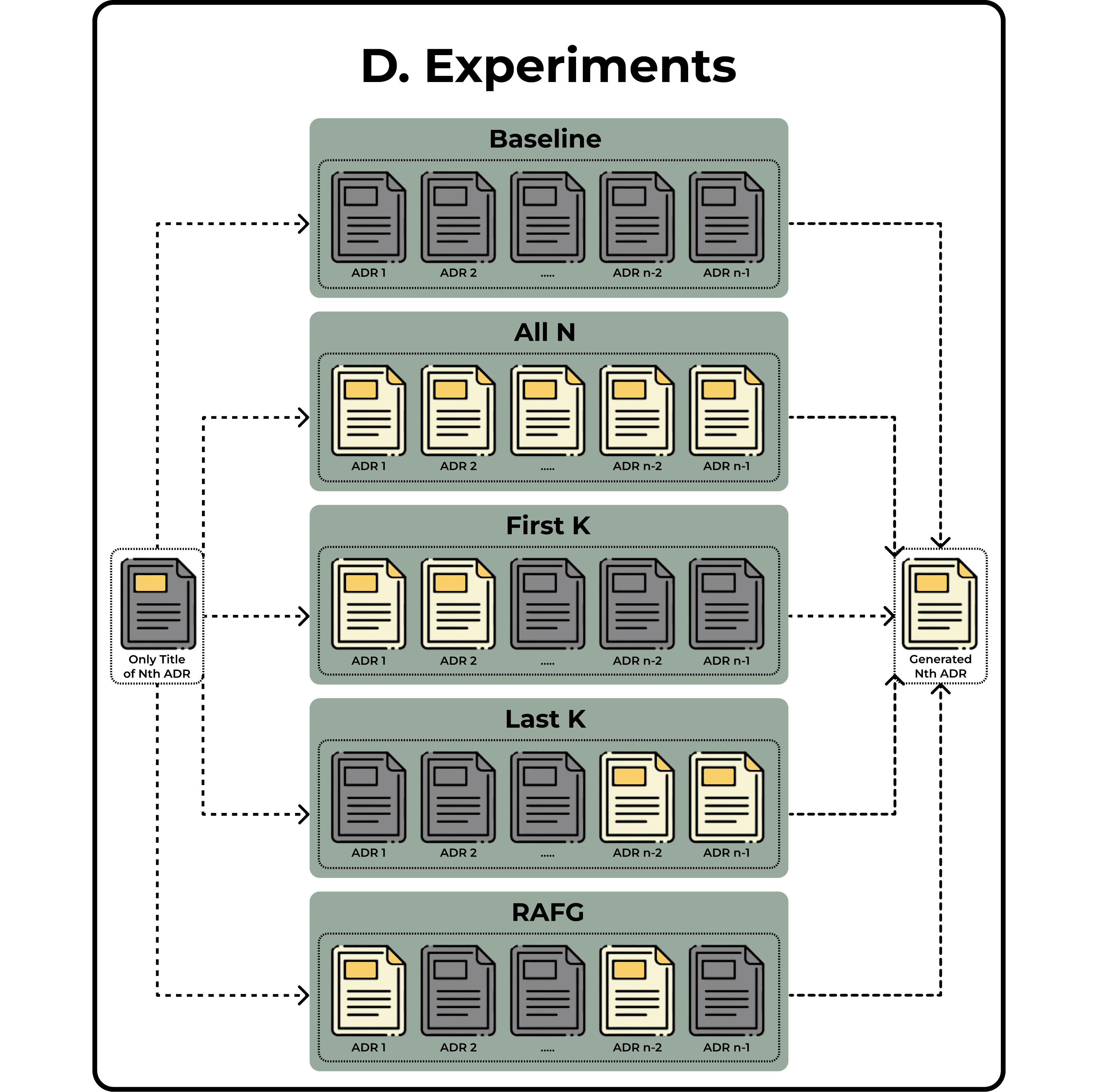}
    \caption{Context Strategies}
    \label{contextStrategies}
\end{figure}

A core hypothesis of this study is that context, the selection and structuring of prior ADRs, plays a critical role in shaping the quality of LLM-generated ADRs. We therefore implemented four distinct context strategies and compared them across a no-context baseline approach. We explain the strategies below whereas Figure \ref{contextStrategies} gives a visual representation.

\textbf{Baseline:} A no-context baseline in which the model is provided only the target ADR title with an instruction to generate the ADR. No prior ADRs from the repository are included.

\textbf{\strategy{All\_N}:} Grounded in literature that defines decisions as being informed by the set of all past decisions, this strategy includes every preceding ADR in the context when generating the next one. It offers the most complete view of a project's decision history and preserves the full evolution of architectural reasoning. However, it may introduce noise from redundant or marginally relevant information and become impractical as repositories grow due to token limitations. While effective for small, well-structured ADR collections, its scalability constraints motivate the exploration of more selective approaches.

\textbf{\strategy{First\_K}:} Drawing on the idea that ADRs capture architecturally significant decisions, many of which are made early in a project, this strategy uses the first $K$ ADRs to anchor generation in foundational design choices. These early decisions often set long-term architectural direction and constraints. However, relying exclusively on early context can overlook recent architectural shifts, evolving system understanding, or changes in technology, limiting its effectiveness in projects with substantial evolution.

\textbf{\strategy{Last\_K}:} To mitigate the limitations of \strategy{First\_K}, this strategy focuses on the most recent $K$ ADRs, capturing the latest architectural developments and ensuring relevance to current design conditions. This recency emphasis can be advantageous, as newer decisions can provide the most pertinent context for subsequent ones. However, they may not always include the background knowledge necessary to fully inform the next decision.

\textbf{Retrieval-Augmented Few-shot Generation (\strategy{RAFG}):} This approach combines few-shot prompting and RAG\cite{izacard2022rafg}, where instead of using a static set of exemplars, the $K$ most contextually similar ADRs are dynamically retrieved from the project history. We utilize the SentenceTransformer model (all-MiniLM-L6-v2) to generate dense embeddings for the query (ADR title) and candidate content (previous ADRs), with cosine similarity used to select the top-$K$ most relevant decisions. This method reduces noise by focusing on semantically related decisions, making it effective for large repositories. However, unlike structured retrieval strategies, it considers the full ADR content rather than chunked sections (e.g., Context, Decision, or Consequences), which may limit the precision of context retrieval. Additionally, there is no external vector database; embeddings and similarity calculations are done in-memory. A key trade-off is the potential disruption of temporal continuity: while retrieval optimizes for semantic overlap, it may bypass the chronological order of decisions, risking the loss of causal relationships inherent in the architectural evolution of the project.

Note that the selection of `$K$' was determined empirically through preliminary experiments, identifying the minimum threshold beyond which the inclusion of additional ADRs failed to yield significant performance gains. This optimization ensures a balance between sufficient contextual depth and model efficiency. Furthermore, while $K$ may fluctuate based on specific repository characteristics or organizational use cases, such hyperparameter tuning is highly context-dependent and falls outside the primary scope of this investigation.

\subsection{Metrics}\label{subsec:metrics}\
We assess generated ADRs against the human-written ground truth using both automated and human-centered evaluation:

\textbf{BERTScore} computes semantic similarity between generated and reference ADRs based on contextualized embeddings from pretrained BERT models. Unlike surface-level metrics, BERTScore captures meaning and conceptual alignment even when wording differs. This makes it particularly suitable for evaluating architectural documentation where paraphrasing is common.~\cite{BERT}.

\textbf{BLEU} measures n-gram precision to capture surface-level overlap between generated and reference texts. Originally developed for machine translation, BLEU penalizes outputs that deviate lexically from the reference, making it sensitive to exact phrasing. While strict, it provides a complementary view of how closely the generated text matches the original formulation.~\cite{BLEU}.

\textbf{ROUGE} provides recall-oriented measures of content overlap, particularly useful for assessing how well key information is retained. ROUGE-1 (unigram overlap) evaluates whether important terms and concepts from the ground truth appear in the generated output. This metric is especially valuable for ensuring that critical architectural details are not omitted. Additionally, ROUGE-L measures the Longest Common Subsequence (LCS), capturing sentence-level structural similarity and fluency without requiring consecutive word matches.~\cite{ROUGE}.

\textbf{METEOR} evaluates alignment between generated and reference ADRs using synonym matching, stemming, and recall-weighted scoring. METEOR goes beyond exact token matching by recognizing morphological variants and semantic equivalents, offering a more flexible assessment than BLEU. Its balanced approach makes it effective for capturing both precision and completeness in generated architectural documentation.~\cite{METEOR}.


\vspace{-2pt}
\section{Results}\label{sec:results}

\begin{table*}[ht]
\centering
\scriptsize
\resizebox{0.92\textwidth}{!}{%
\begin{tabular}{l | l | c c c | c | c | c c | c c}
\toprule
\multirow{2}{*}{\textbf{Model}} & \multirow{2}{*}{\textbf{Experiment}} & \multicolumn{3}{c|}{\textbf{BERTScore}} & \multirow{2}{*}{\textbf{BLEU}} & \multirow{2}{*}{\textbf{METEOR}} & \multicolumn{2}{c|}{\textbf{ROUGE}} & \multicolumn{2}{c}{\textbf{Token Use}} \\
\cmidrule(lr){3-5} \cmidrule(lr){8-9} \cmidrule(lr){10-11}
& & \textbf{F1} & \textbf{Precision} & \textbf{Recall} & & & \textbf{1} & \textbf{L} & \textbf{Avg.} & \textbf{Std. Dev.} \\
\midrule
\multirow{8}{*}{Gemini-2.5}
 & Baseline              & \cellcolor{worst}0.8199 & \cellcolor{worst}0.8267 & \cellcolor{worst}0.8136 & \cellcolor{worst}0.0193 & \cellcolor{worst}0.1433 & \cellcolor{worst}0.2584 & \cellcolor{worst}0.1182 & \cellcolor{worst}1123.41 & \cellcolor{best}188.95 \\
 & All\_N                & 0.8421 & 0.8494 & 0.8354 & 0.0755 & 0.2303 & 0.3593 & 0.1820 & 709.89 & 350.45 \\
 & First\_K (3)          & 0.8401 & 0.8462 & 0.8346 & 0.0746 & 0.2363 & 0.3478 & 0.1778 & \cellcolor{best}636.62 & 305.45 \\
 & First\_K (5)          & 0.8397 & 0.8458 & 0.8340 & 0.0712 & 0.2308 & 0.3502 & 0.1753 & 689.08 & 309.61 \\
 & Last\_K (3)           & \cellcolor{best}0.8440 & \cellcolor{best}0.8505 & \cellcolor{best}0.8380 & \cellcolor{best}0.0866 & \cellcolor{best}0.2428 & 0.3633 & \cellcolor{best}0.1891 & 680.23 & 358.32 \\
 & Last\_K (5)           & 0.8429 & 0.8495 & 0.8367 & 0.0815 & 0.2368 & 0.3632 & 0.1850 & 731.08 & 360.25 \\
 & RAFG (3)              & 0.8438 & 0.8502 & \cellcolor{best}0.8380 & 0.0847 & 0.2423 & 0.3616 & 0.1877 & 674.39 & 358.89 \\
 & RAFG (5)              & 0.8430 & 0.8496 & 0.8368 & 0.0816 & 0.2384 & \cellcolor{best}0.3633 & 0.1858 & 721.69 & \cellcolor{worst}362.34 \\
\midrule
\multirow{8}{*}{Qwen3-235b}
 & Baseline              & \cellcolor{worst}0.8184 & \cellcolor{worst}0.8247 & \cellcolor{worst}0.8126 & \cellcolor{worst}0.0325 & \cellcolor{worst}0.1733 & \cellcolor{worst}0.2626 & \cellcolor{worst}0.1253 & 680.45 & \cellcolor{best}101.03 \\
 & All\_N                & \cellcolor{best}0.8379 & 0.8473 & \cellcolor{best}0.8292 & \cellcolor{best}0.0677 & \cellcolor{best}0.2142 & \cellcolor{best}0.3355 & \cellcolor{best}0.1667 & \cellcolor{best}678.00 & 300.07 \\
 & First\_K (3)          & 0.8344 & 0.8438 & 0.8257 & 0.0608 & 0.2088 & 0.3157 & 0.1577 & 678.98 & 281.59 \\
 & First\_K (5)          & 0.8348 & 0.8436 & 0.8266 & 0.0615 & 0.2098 & 0.3220 & 0.1576 & 735.33 & 306.34 \\
 & Last\_K (3)           & 0.8375 & \cellcolor{best}0.8473 & 0.8283 & 0.0670 & 0.2113 & 0.3275 & 0.1648 & 738.44 & 338.39 \\
 & Last\_K (5)           & 0.8373 & 0.8466 & 0.8286 & 0.0670 & 0.2126 & 0.3321 & 0.1651 & \cellcolor{worst}776.83 & \cellcolor{worst}346.35 \\
 & RAFG (3)              & 0.8375 & \cellcolor{best}0.8474 & 0.8284 & 0.0669 & 0.2118 & 0.3281 & 0.1654 & 727.16 & 333.31 \\
 & RAFG (5)              & 0.8377 & 0.8470 & 0.8291 & 0.0671 & 0.2132 & 0.3333 & 0.1654 & 772.42 & 339.62 \\
\midrule
\multirow{8}{*}{Gemma3-4b}
 & Baseline              & \cellcolor{worst}0.8152 & \cellcolor{worst}0.8182 & \cellcolor{worst}0.8125 & \cellcolor{worst}0.0245 & \cellcolor{worst}0.1630 & \cellcolor{worst}0.2584 & \cellcolor{worst}0.1264 & \cellcolor{worst}814.99 & \cellcolor{best}87.33 \\
 & All\_N                & \cellcolor{best}0.8326 & \cellcolor{best}0.8372 & \cellcolor{best}0.8285 & 0.0651 & 0.2136 & \cellcolor{best}0.3132 & \cellcolor{best}0.1658 & 666.28 & 297.18 \\
 & First\_K (3)          & 0.8265 & 0.8314 & 0.8221 & 0.0528 & 0.2039 & 0.2908 & 0.1510 & \cellcolor{best}610.57 & 173.82 \\
 & First\_K (5)          & 0.8277 & 0.8314 & 0.8245 & 0.0571 & 0.2103 & 0.2988 & 0.1549 & 643.88 & 279.74 \\
 & Last\_K (3)           & 0.8313 & 0.8371 & 0.8259 & 0.0617 & 0.2085 & 0.3044 & 0.1615 & 648.14 & 244.47 \\
 & Last\_K (5)           & 0.8321 & 0.8367 & 0.8280 & \cellcolor{best}0.0654 & \cellcolor{best}0.2160 & 0.3117 & 0.1652 & 665.85 & 312.33 \\
 & RAFG (3)              & 0.8310 & 0.8369 & 0.8256 & 0.0593 & 0.2071 & 0.3031 & 0.1599 & 655.87 & 261.75 \\
 & RAFG (5)              & 0.8315 & 0.8364 & 0.8272 & 0.0629 & 0.2132 & 0.3089 & 0.1621 & 680.16 & \cellcolor{worst}342.95 \\
\midrule
\multirow{8}{*}{GLM-4.6}
 & Baseline              & \cellcolor{worst}0.8186 & \cellcolor{worst}0.8250 & \cellcolor{worst}0.8126 & \cellcolor{worst}0.0275 & \cellcolor{worst}0.1675 & \cellcolor{worst}0.2635 & \cellcolor{worst}0.1256 & \cellcolor{worst}841.88 & \cellcolor{best}292.17 \\
 & All\_N                & 0.8379 & 0.8458 & 0.8307 & 0.0779 & 0.2300 & 0.3327 & 0.1737 & 669.46 & 356.52 \\
 & First\_K (3)          & 0.8365 & 0.8427 & 0.8308 & 0.0755 & 0.2340 & 0.3224 & 0.1699 & \cellcolor{best}579.64 & 307.69 \\
 & First\_K (5)          & 0.8361 & 0.8423 & 0.8304 & 0.0747 & 0.2317 & 0.3240 & 0.1684 & 623.24 & 319.30 \\
 & Last\_K (3)           & 0.8395 & \cellcolor{best}0.8466 & 0.8329 & 0.0839 & 0.2365 & 0.3344 & 0.1781 & 640.84 & \cellcolor{worst}375.30 \\
 & Last\_K (5)           & 0.8394 & 0.8451 & \cellcolor{best}0.8341 & \cellcolor{best}0.0871 & \cellcolor{best}0.2471 & \cellcolor{best}0.3408 & \cellcolor{best}0.1805 & 618.28 & 326.74 \\
 & RAFG (3)              & \cellcolor{best}0.8395 & 0.8464 & 0.8332 & 0.0841 & 0.2383 & 0.3355 & 0.1794 & 620.10 & 349.19 \\
 & RAFG (5)              & 0.8387 & 0.8455 & 0.8323 & 0.0818 & 0.2358 & \cellcolor{best}0.3369 & 0.1766 & 661.47 & 345.25 \\
\bottomrule
\end{tabular}%
}
\caption{Comparative evaluation of automated ADR generation using different LLMs and context strategies. Ground truth ADRs have an average token count of 526.74 (std. dev. = 741.88).}
\label{tab:result}
\end{table*}

\begin{figure*}
    \centering
    \includegraphics[width=0.9\linewidth]{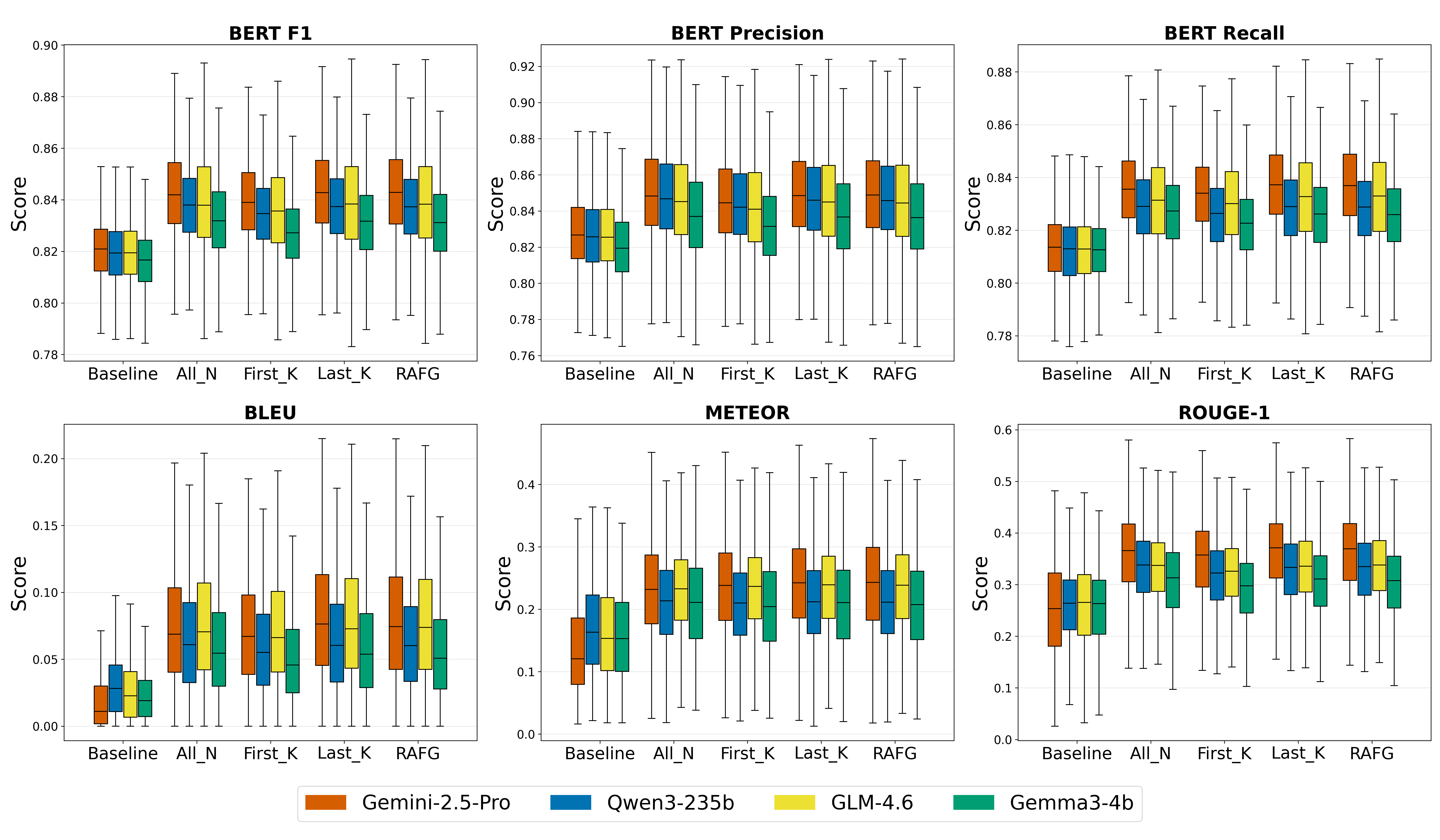}
    \caption{Comparison of model performance across the context strategies (K=3)}
    \label{fig:boxPlot}
\end{figure*}

\begin{figure}
    \centering
    \includegraphics[width=1\linewidth]{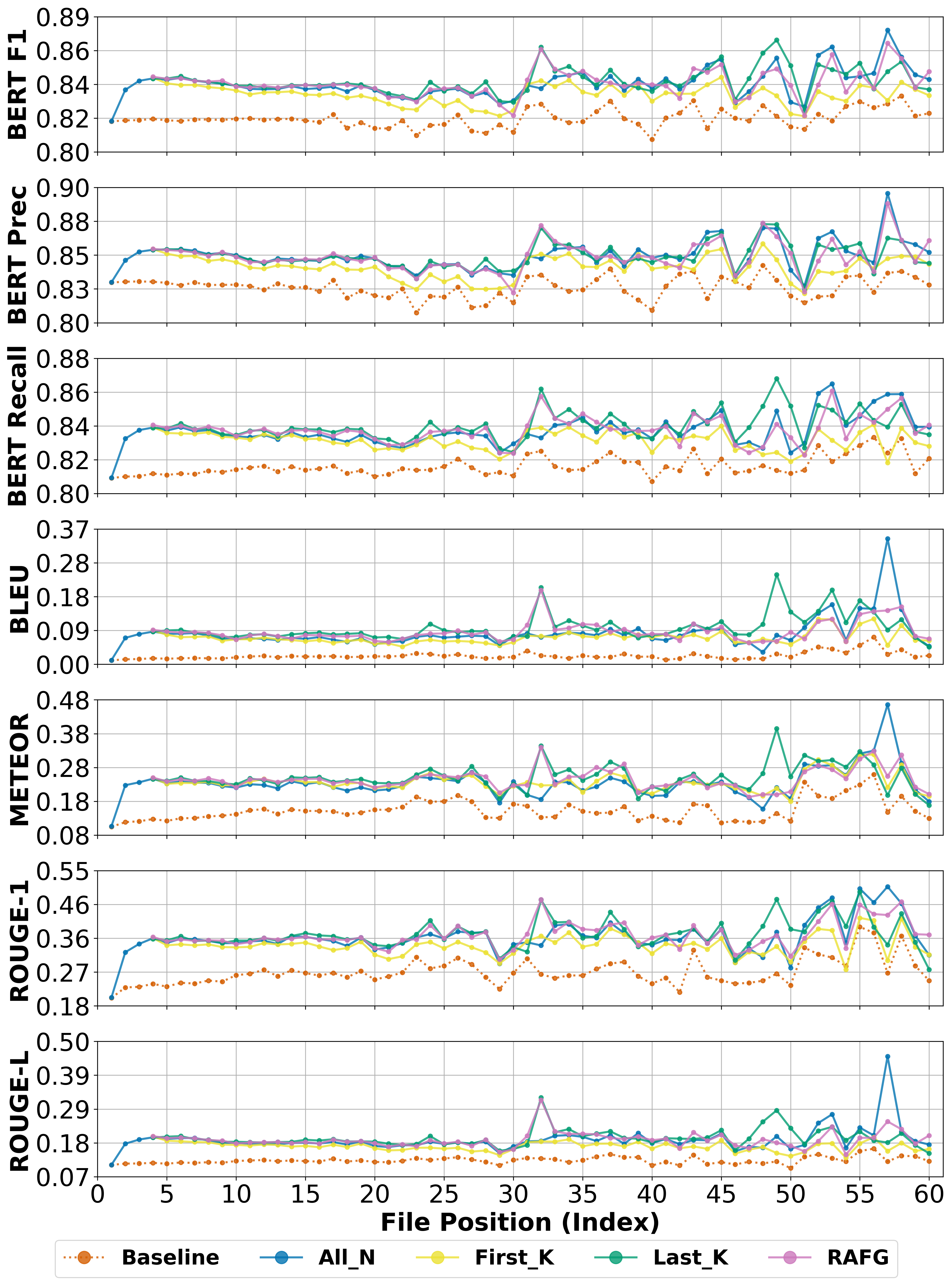}
    \caption{Longitudinal analysis of generation fidelity across ADR positions (K=3). The x-axis represents the sequence position of ADRs within their repository; the y-axis shows the average metric across all ADRs at each position aggregated from all repositories.}
    \label{fig:linePlot}
\end{figure}

The empirical evaluation of context strategies for automated ADR generation reveals a clear relationship between historical context provision and output quality. Our analysis suggests that while \textbf{context is essential}, its selection and volume are subject to significant optimization constraints.

Table \ref{tab:result} presents the average metric values for the various context strategies across models, along with the token usage statistics. Figure~\ref{fig:boxPlot} depicts these values in boxplots to visualize the performance across strategies for each model. In addition, Figure~\ref{fig:linePlot} presents the longitudinal analysis of ADR generation quality relative to the chronological sequence position of the record in its repository.

Supplementing the automated metrics, the authors manually reviewed a subset of the generated samples, focusing on both the content and formatting of the generated ADRs, to understand the underlying differences and shortcomings in the generations that automated metrics alone may not fully capture.

\subsection{Impact of Decision History on Generation Quality}
As illustrated in Figure~\ref{fig:linePlot}, ADR generation quality improves significantly over the baseline when historical context is incorporated (e.g., see `All\_N'), with outputs aligning more closely with human-authored ground truth across all metrics. Table~\ref{tab:result} demonstrates this trend with Gemini-2.5-Pro showing noticeable improvement over the baseline (from 0.0193 to 0.0755 BLEU score under the \strategy{All\_N} strategy). Across all evaluated models and approaches, the no-context baseline consistently emerges as the worst-performing strategy.

The distributional characteristics visualized in the boxplots of Figure~\ref{fig:boxPlot} further corroborate these findings. While the no-context baseline exhibits low variance, it converges on a uniformly suboptimal performance level. In contrast, context-aware strategies demonstrate broader interquartile ranges, indicating that while performance becomes more sensitive to the specific historical context provided, the overall median and peak scores significantly surpass those of the baseline.

Table~\ref{tab:result} also highlights a discrepancy in output length. While human-authored ground truth ADRs maintain a concise average token count of approximately 526.74, the zero-context Baseline for Gemini-2.5-Pro produced a significantly more verbose average of 1123.41 tokens. In contrast, context-grounded approaches like \strategy{All\_N} produced more focused outputs with an average length of 709.89 tokens. This suggests that without historical grounding, LLMs tend to generate extensive, verbose text that fails to replicate the intentionally concise and rationale-heavy nature of professional architectural documentation. Additionally, the standard deviation of generations' token use increases from 188.95 in the baseline to 350.45 (closer to the ground truth value of 741.88), suggesting that context‑aware strategies promote a distributional variance in length that is more representative of human‑authored records, whereas the baseline remains relatively uniform and unspecialized in its scale.

\begin{tcolorbox}[colback={highlight}, colframe={border}]
\small
    \textbf{Answer to RQ1:} Incorporating architectural decision history significantly improves generation quality and reduces excessive verbosity, aligning the output length and rationale with human-authored ground truth.
\end{tcolorbox}

\subsection{Efficacy of Selective Context Strategies}
The longitudinal analysis from Figure~\ref{fig:linePlot} reveals that generation quality improvements typically plateau once a moderate context size is reached. Our results indicate that a limited window of 3 to 5 records typically provides the optimal balance between generation fidelity and computational efficiency. For example, the performance of Qwen3-235b (Table~\ref{tab:result}) reaches a BERTScore F1 of 0.8375 with \strategy{Last\_K} (3), 0.8373 with \strategy{Last\_K} (5), and 0.8379 with \strategy{All\_N} (i.e., all history). In other words, incorporating the complete decision history in context offers negligible benefits compared to a strategically selected subset of decision history.

While \strategy{First\_K} outperforms the baseline consistently (Figure~\ref{fig:linePlot}), it underperforms among the context-grounded approaches across all models, which may indicate that foundational early decisions are less pertinent to current generation tasks. As shown in Table~\ref{tab:result}, Gemini-2.5-Pro using \strategy{First\_K} (3) scores 0.3478 in ROUGE-1 score, falling behind all \strategy{All\_N} (0.3593), \strategy{Last\_K} (0.3633), and \strategy{RAFG} (0.3616) approaches.

The most effective balance of performance and efficiency is found in the \strategy{Last\_K} strategy. This approach demonstrates the strongest overall performance, with Gemini-2.5-Pro reaching its peak BERTScore F1 of 0.8440 using \strategy{Last\_K} (3). This strategy achieves a high effectiveness-to-complexity ratio, providing the model with the most temporally relevant records without the noise of distant history, corroborating that as a project evolves, the immediate architectural context becomes more critical.

Data from Table~\ref{tab:result} also reveals that selective approaches yield output token lengths similar to the exhaustive \strategy{All\_N} approach, further suggesting that complete history is not necessary for effective ADR generation. In practice, the \strategy{All\_N} strategy is highly inefficient, as token usage and associated costs scale linearly with repository size. This makes the exhaustive approaches impractical for long-lived projects, whereas selective grounding maintains performance while remaining scalable for extensive decision histories.

\begin{tcolorbox}[colback={highlight}, colframe={border}]
\small
    \textbf{Answer to RQ2:} Selective context strategies achieve performance comparable to exhaustive approaches, suggesting that the complete architectural history is not necessary for effective ADR generation. Notably, \strategy{Last\_K} consistently outperforms \strategy{First\_K} and often matches or exceeds \strategy{All\_N} fidelity, indicating that the recent architectural trajectory is a more potent predictor of rationale than foundational history.
\end{tcolorbox}

\subsection{Comparison of Chronological and Retrieval-Based Context}
Architectural scenarios often involve cross-cutting concerns or non-sequential dependencies that often rely on context that is temporally distant yet conceptually relevant. In these instances, the rigid chronological windows of \strategy{Last\_K} and \strategy{First\_K} strategies may prove insufficient as they can bypass critical foundational decisions that remain pertinent to current changes. By employing the retrieval-based \strategy{RAFG} approach, models can dynamically surface these relevant but distant precedents, potentially mitigating architectural drift.

Data in Table~\ref{tab:result} indicates that the \strategy{RAFG} strategy consistently outperforms \strategy{First\_K} and achieves performance parity with \strategy{Last\_K}. For example, the Qwen3-235b model produces a METEOR score of 0.2088 under the \strategy{First\_K} (3) strategy, which increases to 0.2113 for \strategy{Last\_K} (3) and 0.2118 for \strategy{RAFG}. This competitive convergence is further evidenced in the longitudinal analysis of Figure~\ref{fig:linePlot}, where the trajectories for \strategy{All\_N}, \strategy{RAFG}, and \strategy{Last\_K} remain tightly clustered. Their interwoven performance throughout the project timeline suggests that these strategies offer comparable utility in capturing evolving architectural rationale.

Upon inspecting a sample of the outputs, we noticed that \strategy{RAFG} seem to outperform the chronological strategies by addressing scenarios defined by conceptual locality rather than simple temporal proximity. While aggregate metrics show parity with \strategy{Last\_K}, we found that \strategy{RAFG} may provide superior utility for decisions involving cross-cutting concerns or non-sequential dependencies, where the pertinent architectural context is often temporally distant.

\begin{tcolorbox}[colback={highlight}, colframe={border}]
\small
    \textbf{Answer to RQ3:} While \strategy{RAFG} and chronological \strategy{Last\_K} demonstrate aggregate performance parity, \strategy{RAFG} offers a specialized advantage for non-sequential or cross-cutting architectural scenarios by prioritizing conceptual relevance over temporal proximity, albeit at the expense of increased implementation complexity and retrieval latency.
\end{tcolorbox}



\vspace{-2pt}
\section{Discussion}\label{sec:discussion}

\subsection{Interpretation of Results}
The results of this study provide an understanding of the interplay between architectural context and LLM performance in automated documentation tasks. Our primary findings of context selection strategies showing a more pronounced influence on generation quality than model scale may drive a shift for automated AKM. Our results support a growing thread of studies showing that in certain specialized tasks (in this case, ADR generation), the bottleneck is not the reasoning capacity of the LLMs, but the strategic context provided to the prompt.

The comparative performance of context strategies reveals two fundamental principles of architectural decision-making: temporal locality and conceptual locality. The \strategy{Last\_K} strategy, using only 3--5 recent records, achieves results comparable to the \strategy{All\_N} strategy, which incorporates the complete decision history. This performance parity suggests that architectural decisions regularly have strong temporal locality, i.e., most of the relevant forces for a new decision (e.g., constraints and trade-offs) are established by recent decisions rather than older ones. From an architectural perspective, this finding aligns with the principle that system evolution is dominated by local modifications rather than wholesale restructuring. While foundational decisions establish the system structure, subsequent decisions form evolutionary trajectories constrained primarily by their immediate predecessors.

However, temporal locality does not characterize all architectural decisions uniformly. The \strategy{RAFG} strategy demonstrates superior performance for cross-cutting concerns that span multiple system components or reactivate dormant architectural patterns. These decisions show conceptual locality, where semantic relationships outweigh temporal proximity. For instance, a decision regarding authentication mechanisms may draw more from a security decision made months prior than from recent database schema modifications. This distinction suggests that effective context selection must account for the architectural scope of the decision being documented.

After reviewing a sample of ADR generations across some repositories, we found some evidence suggesting patterns attributable to documentation practices rather than model limitations. Most observed performance degradation was associated with two prevalent documentation issues: external content dependency and knowledge vaporization. In the former, ADR creators replace architectural rationale with references to external resources such as wiki pages, design documents, or issue trackers. In the latter, ADRs rely on implicit organizational knowledge that exists only in institutional memory. While these factors may help explain certain trends observed in the review, establishing causality would require deeper analysis, which falls outside the scope of this paper. 

These practices create a fundamental evaluation challenge. When ground-truth ADRs link to external content, LLMs generate substantive, self-contained documentation that our metrics penalize even if the LLM-generated ADR is architecturally valid. The model produces documentation that would serve future maintainers, yet scores poorly against reference text that assumes access to resources unavailable in the repository. This misalignment represents not a failure of automated generation but a measurement artifact arising from inadequate ground truth. This phenomenon also reveals how current documentation practices transfer maintenance burden from documentation to human memory, creating barriers for both automated tools and human developers unfamiliar with organizational context.
    

In light of this, we posit that LLM-based generation can help surface gaps in how architectural rationale is documented. Performance drops in these cases may reflect missing or implicit context in the reference ADRs rather than generation errors alone. When an ADR relies on undocumented assumptions or external discussions, generating a coherent and aligned record becomes difficult in much the same way it can hinder onboarding for new contributors. While not a primary focus of this study, these observations indicate that integrating LLMs into ADR workflows may encourage more explicit and self-contained documentation practices, which could improve the clarity and durability of architectural knowledge over time.

\subsection{Implications for Practitioners and Researchers}

\subsubsection{\textbf{Practitioners}}

Organizations implementing automated ADR generation should consider three strategic factors based on our findings. First, recency-based context selection provides a pragmatic default strategy. The \strategy{Last\_K} strategy achieves comparable performance to more sophisticated retrieval methods while eliminating infrastructure requirements and retrieval latency. This simplicity reduces implementation barriers and operational complexity, making automated documentation accessible to organizations without dedicated AKM infrastructure.

Second, model scale presents a less critical constraint than previously assumed for architectural documentation tasks. Compact models such as Gemma3-4b, when provided with appropriate context, generate ADRs comparable in quality to models with order-of-magnitude more parameters. Similarly, architectural innovations such as graph augmentation (GLM-4.6) achieve competitive quality while demonstrating superior token efficiency, suggesting that model design choices beyond parameter count significantly influence practical utility. For organizations with data governance requirements that preclude cloud-based API usage, locally deployable models offer a viable path to automation without accepting substantial quality degradation. This finding reduces both computational costs and data privacy concerns associated with architectural knowledge management automation.

Third, maximizing the effectiveness of automated tools requires maintaining self-contained architectural documentation. The performance degradation associated with external references demonstrates how incomplete documentation penalizes both automated generation and human understanding. Organizations should establish practices that capture architectural rationale directly within ADRs rather than deferring to external resources. This investment improves not only automated tool performance but also long-term documentation utility as external dependencies decay or become inaccessible. Additionally, automated generation can retrospectively fill gaps in incomplete ADR collections, helping organize recovery efforts for undocumented architectural decisions.

\subsubsection{\textbf{Researchers}}

Several research opportunities emerge from our findings. Current context selection strategies treat architectural decisions as uniformly dependent on either temporal or semantic proximity. However, our results suggest that decision type influences optimal context selection. Future work should develop classification methods that identify decision characteristics, such as scope, architectural layer, or novelty, and adapt context selection accordingly. Hybrid strategies that combine recency for incremental decisions with semantic retrieval for cross-cutting concerns could outperform uniform approaches.

Beyond predefined selection strategies, exploring arbitrary ADR selection as context presents opportunities for understanding architectural dependencies. Rather than assuming temporal or semantic proximity determines relevance, future work could examine which specific ADRs, regardless of their position in the decision history, contribute most to generating accurate documentation. This approach would reveal implicit architectural dependencies and decision interdependencies that current strategies overlook. Graph-based representations offer particular promise for modeling these relationships, where nodes represent ADRs and edges capture dependencies, influences, or shared architectural concerns. Such structures could expose architectural patterns, identify foundational decisions that propagate constraints through subsequent choices, and detect decision clusters that address related quality attributes.

Furthermore, integrating source code and linking ADRs to the code they seemingly govern creates bidirectional traceability: from decisions to their implementation artifacts and from code back to its architectural rationale. Analyzing these connections over time could reveal when implementations deviate from documented intent, whether ADRs remain synchronized with evolving codebases, and how refactoring efforts propagate through architectural documentation. This holistic view of the decision-code relationship would provide both a diagnostic tool for detecting architectural inconsistencies and a foundation for maintaining alignment between architectural intent and implementation reality.

The ground-truth quality issues observed in our evaluation highlight the need for curated, long-form architectural decision datasets. Current datasets reflect the documentation practices that create barriers to automated generation, including external dependencies and fragmented rationale. Research-grade datasets should prioritize self-contained ADRs with complete architectural justification to enable meaningful evaluation of generation quality independent of documentation debt. Such datasets would also facilitate research on detecting and addressing incomplete documentation.

Finally, the architectural locality principles identified here warrant investigation across different software domains and organizational contexts. Our dataset encompasses diverse projects, but further research should examine how factors such as system maturity, domain, and architectural style influence the temporal and conceptual locality of architectural decisions. Understanding these relationships would enable context selection strategies tailored to specific development contexts, improving both automated generation quality and our understanding of architectural knowledge evolution.

\vspace{-2pt}
\subsection{Threats to Validity}
While this study provides an evaluation of various context strategies for automated ADR generation, several factors may influence the interpretation and generalizability of the findings:

Regarding \textbf{construct validity}, several measurement challenges affect our evaluation. Our metrics consequently measure replication fidelity rather than absolute generation quality; thus higher similarity does not necessarily indicate better documentation, only closer alignment with the (potentially flawed) reference. This discrepancy is exacerbated because LLMs produce self-contained outputs, while human ADRs often rely on implicit or external context (wikis, design documents, code snippets), leading to artificially low similarity scores.
Using only the ADR title as input can be misleading due to title heterogeneity: ambiguous titles (e.g., ``migrating to a new language codebase'') provide less actionable signal than specific ones (e.g., ``switching to Python''), inflating variance in scores. Our no-context baseline represents the realistic first-decision scenario, though alternative baselines (e.g., random-K) might provide complementary insights. Finally, our manual review of generated samples was exploratory rather than systematic—two authors informally examined low and high-scoring outputs without predefined criteria, rubrics, or inter-annotator agreement measures—limiting its validity as a rigorous complement to automated metrics.

Next, looking at \textbf{internal validity}, the automated extraction of ADR titles, used as the primary input for generating subsequent records, relied on internal sub-scripts that exhibited occasional inaccuracies. Although these cases were identified as insignificantly low ($<$0.2\% of the dataset), improper title generation in these instances could disproportionately impact the model's generation performance for those specific samples.

Concerning \textbf{external validity}, architectural documentation practices are highly use-case dependent, and it is possible that specific project domains or organizational cultures may favor context strategies that differ from our general findings. K values were selected via preliminary inspection of diminishing returns in the All-N strategy rather than systematic sensitivity analysis; while this pragmatic approach was sufficient for our research goals, comprehensive K optimization across project types--though outside this work's scope--would strengthen generalizability. Furthermore, LLM generations are inherently probabilistic. While the large scale of our experiments (80,000+ generations) aims to provide a reliable net average for comparison, the results should be interpreted as a general trend rather than a universal rule for every software system.

Finally, focusing on the \textbf{conclusion validity}, since our dataset was sourced from open-source repositories, it is possible that some ADRs were included in the LLMs' training data, potentially inflating absolute performance scores. However, our study focuses on comparing the relative improvement of different context strategies against a baseline approach that uses the same dataset. Any training data overlap affects all strategies equally, making the comparative analysis robust. Additionally, it is natural and expected for LLMs to possess general knowledge about common programming languages, technology stacks, and architectural patterns, which enhances their utility for practical software engineering tasks rather than undermining the validity of our findings. While our model diversity strengthens generalizability, controlled size comparisons within the same architecture would further validate our claim about context mattering more than model scale.

\vspace{-2pt}
\section{Related Work}\label{sec:relatedWork}

Several studies have been conducted over the last few years to understand the extent to which Generative AI can aid software architecting practices. Esposito et al. ~\cite{esposito2025genai} performed a multi-vocal literature review on the application of generative AI in software architecture. The study clearly points out that while most of the work in applying Generative AI to software architecture has been focused on decision support, more work needs to be done on the contextual aspects. Particularly with respect to architecture design, Dhar et al.~\cite{dhar2024leveraging} proposed an approach that uses a combination of data extraction and retrievals using different types of GenAI to better manage architecture knowledge. There have also been works that focused specifically on architecture decision records. Dhar et al.~\cite{dhar2024can} investigated whether LLMs could act as architects. They evaluated different types of models on their ability to generate the "Decision" section of an ADR given a "Context". Their results demonstrated that while state-of-the-art models can generate relevant decisions in a zero-shot setting, they often lack domain-specific nuance. Crucially, they found that smaller, cost-effective models (e.g., GPT-3.5) could achieve performance comparable to larger models when utilizing few-shot prompting strategies. This finding highlights the importance of \textit{context engineering} over raw model size. Further, Diaz-Pace et al.~\cite{diaz2024helping} proposed a design assistant approach that improves the reflective practices of novice architects and enables them to improve design decisions captured using ADRs. 
Extending the work, Diaz-Pace et al. proposed an agentic approach for improving the architectural design decisions by doing analysis between different alternatives. The work was evaluated in a case study that involved deciding between different patterns.  Building on all these, Soliman and Keim~\cite{soliman2025do} investigated the depth of architectural knowledge embedded within LLMs, finding that while models possess broad theoretical knowledge, they often struggle to apply it consistently to specific constraints without guidance.  

Apart from architecture knowledge and architecture decision-making, there have also been works that discuss the potential of GenAI in aiding architects. A method that leverages AI, particularly LLMs, to semi-automate the generation of candidate architectures from requirements was proposed by Eisenreich et al.~\cite{eisenreich2024}. Jahić et al.~\cite{jahicsaml} conducted a study to understand the capabilities, challenges and limitations of LLMs in practice with respect to software design. Fuchß et al.~\cite{fuch2025} developed an approach that leverages LLMs for extracting component names as software architecture models from source code and software architecture documentation to enhance traceability. Arun et al.~\cite{arun2025llms} explored the generation of structural architectural components, specifically within serverless domains. Their findings suggest that while LLMs can generate valid architectural structures, the "hallucination" of non-existent dependencies remains a challenge, further motivating the need for robust context retrieval mechanisms to ground the model's generation in the reality of the existing codebase. 

Departing from the aforementioned approaches that leverage LLMs or, in general, Generative AI for software architecture knowledge management or generating design decisions, as well as building on the contextual challenges highlighted by several existing studies~\cite{esposito2025genai,dhar2024can}, in this work, to the best of our knowledge, we have performed a first-of-its-kind extensive study to further investigate the amount of context required and its impact in generating architecture decision records using large language models. 



\vspace{-2pt}
\section{Conclusion and Future Work}\label{sec:conclusions}

We presented a large-scale empirical study evaluating how different context-provision strategies affect automated ADR generation using LLMs. Our results demonstrate that providing contextual ADR history substantially improves generation quality over a no-context baseline, while a small recency window (\strategy{Last\_K}, typically 3 to 5 records) yields near-optimal performance with favorable efficiency. Furthermore, we found that simple recency heuristics often match more complex retrieval approaches (\strategy{RAFG}) in aggregate effectiveness. These findings suggest that once adequately grounded by context, smaller or locally deployable models can produce competitive results, shifting the primary optimization burden from model scale to context selection and engineering. 

In summary, this study highlights that context strategy matters more than model size and that expanding context length past a modest $K$ often yields negligible gains. While \strategy{Last\_K} serves as an effective and low-complexity default for ADR automation, \strategy{RAFG} remains beneficial for capturing cross-cutting or conceptually distant dependencies. Additionally, we observed that data and documentation hygiene (specifically maintaining self-contained ADRs and avoiding external stubs) materially affects both automated generation and evaluation fidelity.

We outline several avenues for future work to extend and strengthen these findings. This includes developing hybrid retrieval strategies that combine recency heuristics with semantic retrieval to better handle architectural drift. We also propose incorporating multi-artifact grounding, such as code and design diagrams, to better align ADR generation with implementation reality. Future efforts should also focus on improved evaluation protocols that capture ADR utility beyond text-similarity, dataset expansion to improve reproducibility, and longitudinal deployment studies to evaluate developer acceptance and workflow integration in realistic settings.

\vspace{-2pt}
\section{Data Availability}\label{sec:data}

All experimental data, scripts, and results are available in our replication package: \url{https://zenodo.org/records/18370195}.

\bibliographystyle{unsrtnat} 
\bibliography{references.bib}

\end{document}